# The mass-energy equivalence and length contraction are consistent with compression of a fluidic aether with a speed of sound equal to the speed of light

Johannes D. Johansson

Department of Biomedical Engineering

Linköping University

581 85 Linköping, Sweden

Phone: +46 (0) 10 – 103 24 64

Fax: +46 (0) 13 – 14 94 03

johannes.johansson@liu.se

johannesj@gmail.com

ORCID: 0000-0003-4910-0291

## Abstract
**The Lorentz aether theory was the mathematically equivalent precursor to the special theory of relativity (SR). It assumed the existence of a static aether filling space in which particles are affected by the Lorentz factor for length contraction to explain the null result of the Michelson-Morley experiment. However, it was early abandoned in favour of SR due to the apparent arbitrary introduction of the Lorentz factor. The aether has thereafter been assumed not to exist as it is seemingly undetectable and not needed for the SR. In this discussion paper it is argued that this reasoning to reject the aether is unfounded since the Lorentz factor also appears as an increased resistance term in fluid dynamics when the speed of sound is approached. This means that, for example, particle accelerator tests of the Lorentz factor can be interpreted as a form of detection of a fluidic aether. Furthermore, it is shown that the mass-energy equivalence is identical to the energy for compression of a fluid with the speed of sound equal to the speed of light. It is thus possible that the aether not only exists but that it also is the matter all particles are made of.**

## 1 Introduction

What is light? This is a question that has been under great debate during the centuries. Christiaan Huygens and Robert Hooke thought that light was a form of waves like sound. Huygens believed these waves propagated through a hypothetical material called the aether which was believed to still exist even in vacuum and to prevent propagation of sound but allow propagation of light [1]. Isaac Newton on the other hand thought light was particles (corpuscles) on the basis that light does not bend much around corners, though he still thought they caused vibrations in the aether [2]. In the early 19$^{th}$ century, Thomas Young showed that light indeed is a wave phenomenon [3] and some years later François Arago and Augustin-Jean Fresnel showed that these waves need to be completely transverse in order to explain the phenomenon of polarization [4]. In the middle of the same century, James Maxwell showed that electromagnetic waves propagate with the same speed as light [5] and hence made the conclusion that light is transverse electromagnetic waves.

George Stokes imagined a fluid-like aether that had a velocity of 0 m/s at planetary surfaces as expected of a fluid. He used this model to describe the phenomenon of aberration of light from stars and planets [6] and the idea of a fluid-like aether with zero velocity at the earth surface was later seemingly supported by the famous experiments of Michelson and Morley [7, 8]. However, when Hippolyte Fizeau earlier had measured the changes in light velocity from travelling through moving light, it looked like the aether wasn't moved by the water molecules and the motion component only depended on the increase in index of refraction of the water compared to free space [9]. George FitzGerald and Hendrik Lorentz came independently up with a solution to this contradiction by introducing the concept of length contraction, where objects are assumed to contract by the Lorentz factor in the direction of travel relative to the aether [10],[11]. However, e.g. Albert Einstein and Hermann Minkowski disliked this explanation as there was no known physical basis to why the length contraction should occur. As Einstein's theory of special relativity could explain the length contraction without any need for an aether to exist, scientists eventually abandoned the aether theory in favour of viewing light as photons with both wave and particle-like characteristics and which do not need any medium for propagation. In this paper, I would like to point out that length contraction as well as the mass-energy equivalency are consistent with compression of a fluidic aether. It should be stressed that it is not meant to question the mathematics of either special or general relativity, though, only their interpretations.

## 2 Compression of Ideal Gases and its Implication on the Mass of Particles

Assume that a small mass, d$m$, (kg) is compressed into a volume, $V$ (m³) in thermal equilibrium with its surrounding (Figure 1). The energy, $E$ (J), for isothermal compression into a gas tank with volume $V$ is given by [12]

$$E = pV \ln\left(\frac{p}{p_0}\right) \quad (1)$$

where $p$ (N/m²) is the pressure in the tank and $p_0$ the pressure in the surrounding. Let us now derivate this expression with respect to mass density, $\rho$, (kg/m³). The speed of sound, $c$, (m/s) is related to changes in pressure and mass density according to

$$c^2 = \frac{dp}{d\rho} \quad (2)$$

This together with equation (2) gives

$$\frac{dE}{d\rho} = \frac{c^2 dE}{dp} = c^2\left(V \ln\left(\frac{p}{p_0}\right) + V\right) = c^2 V\left(\ln\left(\frac{p}{p_0}\right) + 1\right) \quad (3)$$

and with $m = \rho V$ this can be rewritten as

$$dE = d\rho \cdot V c^2 \left(\ln\left(\frac{p}{p_0}\right) + 1\right) = dm \cdot c^2 \left(\ln\left(\frac{p}{p_0}\right) + 1\right). \quad (4)$$

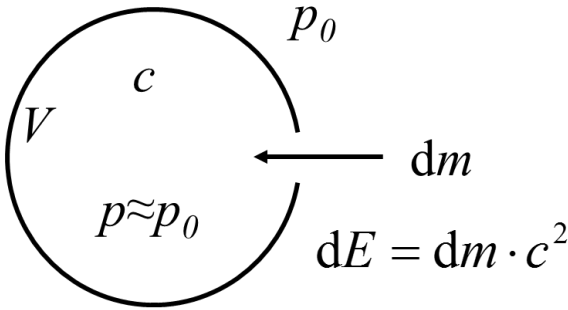

*Figure 1 A small mass dm of a fluid with the speed of sound c is compressed into the volume V. If the amount is small enough so that the pressure change from the surrounding also is small, the energy required for this, i.e. the integral of the pressure change dp, is dE = dm·c².*

If the pressure change relative to the surrounding is small, this becomes

$$dE \approx dm \cdot c^2 \quad (5)$$

i.e., if the pressure change relative to the surrounding is small, the energy, d$E$ (J), to compress the mass d$m$ equals the mass times the speed of sound squared. Now, does this expression not look vaguely familiar? The famous expression

$$E = m \cdot c_0^2 \qquad (6)$$

is usually interpreted as an equivalency between mass and energy, with mass being a form of energy that can be converted into other forms of energy. It was originally expressed by Einstein as "Gibt ein Körper die Energie $L$ in Form von Strahlung ab, so verkleinert sich seine Masse um $L/V^2$" – "If a body emits the energy $L$ in the form of radiation, its mass reduces by $L/V^2$" where "$V$" in this case was used to denote $c_0$ [13], i.e.

$$\mathrm{d}m = \frac{\mathrm{d}E}{c_0^2} \qquad (7)$$

However, in the identical equation (5), there is no conversion or equivalency between mass and energy. It just describes the energy required to compress a certain amount of mass. Is equation (6) describing the same thing, i.e. the mass and rest energy of a mass particle is due to the compression of a fluid with the speed of sound equal to the speed of light in vacuum, $c_0$? That sounds like the hypothetical aether. So why do people no longer believe in it? Modern physics literature usually cites the famous Michelson-Morley experiment as the reason to doubt the existence of the aether but, as mentioned, it actually shows just what is to be expected of a hypothetical fluid aether: The velocity of the aether relative to the earth is 0 m/s at the surface. The problematic experiment is rather the measurements of the Fresnel drag by flowing water made by Hippolyte Fizeau which showed that the motion of the hypothetical aether must be independent of the direction of the flowing water. This is a very odd behaviour of a fluid which would have been expected to follow the water. It could possibly be reconciled with Stokes' view if the aether is expected to move with the earth but not with the water. However, that would still be a very strange behaviour for a fluid.

Another attempt to reconcile the Michelson-Morley and Fizeau experiments was suggested by George FitzGerald [10] and implemented by Hendrik Lorentz in 1895 [11]. It relied on the concept of length contraction by the Lorentz factor

$$\frac{1}{\sqrt{1-v^2/c_0^2}} \qquad (8)$$

where $v$ (m/s) is the speed of the object relative to the aether. This solution solves the contradiction between the Fizeau and Michelson-Morley experiments but was criticized by e.g. Albert Einstein and his former teacher Hermann Minkowski for being *ad hoc* though. Minkowski wrote that "This hypothesis sounds rather fantastical. For the contraction is not to be thought of as a consequence of resistances in the ether, but purely as a gift from above, as a condition accompanying the state of motion." [14]. As Einstein's special theory of relativity from 1905 [15] was seen as a simpler solution to the paradox, it soon superseded FitzGerald's and Lorentz' ideas.

## 3 The Prandtl – Glauert approximation and the Lorentz factor

So how does Minkowski's criticism of the Lorentz aether hold up? Let us move forward to the 1920s and take a look at the field of aerodynamics and more precisely the Prandtl-Glauert approximation [16] which relates the dimensionless pressure coefficient, $C_p$ (-), on an object flying through compressible fluid (Figure 2) to the corresponding pressure coefficient in an ideal incompressible fluid, $C_{p0}$ (-), according to

$$C_p = \frac{C_{p0}}{\sqrt{|1 - M_\infty^2|}} \qquad (9)$$

where $M_\infty$ (-) is the freestream Mach number which relates the velocity of the object, $v$ (m/s), to the speed of sound by

$$M_\infty = \frac{v}{c} \qquad (10)$$

giving

$$C_p = \frac{C_{p0}}{\sqrt{|1 - v^2/c^2|}} \qquad (11)$$

which contains the familiar Lorentz factor, equation (8).

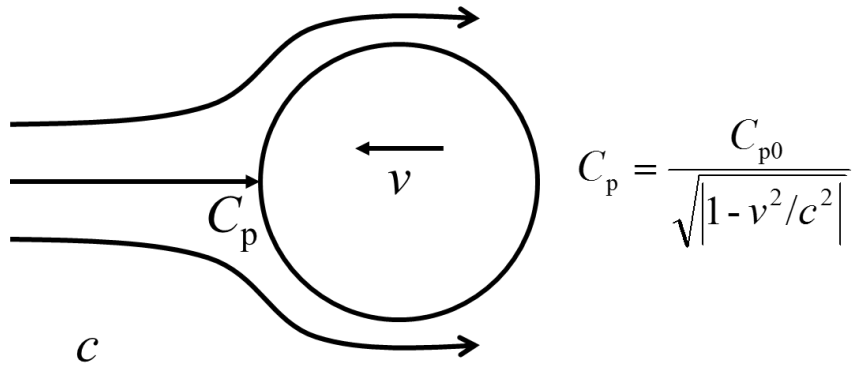

Figure 2 The relation between the pressure coefficient, $C_p$ (-) at the surface of an object travelling with the speed $v$ (m/s) through a compressible fluid with the speed of sound $c$ (m/s) and the pressure coefficient if it had been travelling through an ideal incompressible fluid, $C_{p0}$ (-), is $1/\sqrt{|1-v^2/c^2|}$, i.e. the Lorentz factor.

So, if mass particles would move in a hypothetical fluid aether, they are expected to be exposed to the same increase in resistance. It thus turns out to be exactly the opposite of what Minkowski said: The contraction *is* to be thought of as a consequence of resistances in the aether! If anything, there would have been reason to doubt the existence of the aether if the Lorentz factor *had not* shown up. Sadly, both Minkowski and Lorentz had already died by the time Glauert published the formula in 1928 and thus never got to see the similarity with the fluid mechanics. Lorentz died at a respectable age of 74 years earlier the very same year of 1928 but Minkowski had the misfortune to get a ruptured appendicitis before the introduction of effective antibiotics and died way too young at the age of 44 in 1909. It is an unpleasant reminder of a deadly world we could be returning to due to the emergence of antimicrobial resistance caused by inappropriate use of antibiotics in recent years. Perplexingly, the relation seems to have flown under the radar for almost everyone else too. Rienstra & Hirschberg notes that the Lorentz and Prandtl-Glauert factors are the same [17] but do not mention the aether. One Akira Nishida has recently stated outright that this similarity is consistent with a fluidic aether though [18]. Ironically, Einstein of all people still believed in some sort of aether [19] but his concept of this was very abstract and he never got acceptance for the idea. Paul Dirac also did a likewise abstract argument for the existence of an aether in 1951 [20] but it did not gain much support either. In the end, evidence of the elusive aether showed up just after the hunting party had given up and gone home.

An argument against the similarity between the Lorentz factor and the Prandtl-Glauert factor is that the latter acts like an increased force on the airplane while the former gives an increase in energy to achieve speeds close to the speed of sound and light respectively. It may be possible that this difference is due to that the airplane is an entity distinct from the surrounding air while a mass particle would be a compressed part of the surrounding hypothetical aether. If it is assumed that the aether can be viewed as an ideal gas, the relation between the speed of light, the temperature, $T$ (K), and the mass, $m_a$ (kg) of the individual particles of the aether ("aetherons") is given by

$$c_0 = \sqrt{\frac{\gamma k T}{m_a}} \tag{12}$$

where $\gamma$ (-) is the adiabatic index and $k$ the Boltzmann constant ($1.38 \cdot 10^{-23}$ J/K). If it is assumed that the aether behaves like a diatomic gas with $\gamma = 1.4$ and a temperature of 3 K, we obtain an estimate of the aetheron mass of approximately $6.4 \cdot 10^{-40}$ kg. While these assumptions are rather wild guesses, it nevertheless seems reasonable that such small particles easily could be gained and lost by the mass cloud of e.g. a much larger accelerated electron ($m_e = 9.1 \cdot 10^{-31}$ kg). Further, the apparent mass increase when a particle is accelerated to relativistic speeds may very well then be an actual mass increase from the aether compression in front of the particle.

Another problem with the idea of a fluid aether is that fluids generally do not support transverse waves. However, helium-3 becomes a fluid that loses its viscosity when at sufficiently low temperatures in a state called superfluid. In this state, it has been found to be able to support acoustic transverse waves [21]. As a matter of fact, while Dirac did not get much support for his aether ideas, they did inspire Sinha, Sivaram and Sudarshan to 1975 suggest the superfluid vacuum theory where the aether is assumed to exist as a superfluid [22]. They also suggested the aether to be a suitable candidate for the so-called "dark matter" [23] astrophysicists believe make up a majority of the mass of the universe. ("Dark matter" is somewhat of a misnomer as dark objects absorb light whereas "dark matter" is transparent. E.g. "invisible matter" or "transparent matter" would have been a better name.) At the time, they found much better agreement with the mass density estimate of the universe if such an aether component was added to the visible mass. My personal speculation is that mass particles are tiny hurricane- or galaxy-like vortices in the aether that for fortunate but unknown reasons stay stable in certain configurations. As a matter of fact, William

Thomson (Lord Kelvin) had already in 1867 suggested that atoms are vortices in a viscousless (i.e. superfluid) aether [24]. It can be noted that if this is true, Descartes may have been on the right path when suggesting that gravity is due to vortices in the aether even though he lacked models to develop that idea into a useful theory [25]. It would be interesting to investigate whether the compression of the aether in such a vortex could be matched with the theory of general relativity, which describes gravity. It can also be noted that wave-particle duality becomes quite intuitive for such vortices as the wave properties would be associated with their spinning.

The Prandtl-Glauert approximation is only valid up to Mach numbers of about 0.7, so the Lorentz factor could also have been expected to be limited in its region of validity. However, it has been found to be valid much closer to the singularity point, at least for acceleration of electrons in particle accelerators [26]. Still, faster-than-light travel might be possible for an object sufficiently far away from any source of gravity in space if this allows it to drag the aether along with it as in the Stokes model. If Lorentz' hypothesis of a universal stationary aether is correct, the prospect of breaching the speed of light looks practically impossible though. Also, from experimental proofs of time dilation such as the Hafele-Keating experiment [27], it seems that a hypothetical aether does not rotate with the earth as a clock in a plane flying eastwards with the earth's rotation measures a shorter time than a clock stationary on earth's surface while a clock flying westwards against the earth's rotation measures a longer time.

Fun fact: In the same paper as he presents his idea of light as particles, Isaac Newton speculated that all matter is condensation of the aether [2] as would be the case if Kelvin's hypothesis of atoms being vortices in the aether is correct. Incidentally, Newton was also the first to come up with an analytical expression for the speed of sound as a function of pressure and density, although it was of course impossible to know at this time that it would support his hypothesis about aether compression since the apparent mass-energy equivalence was discovered much later.

## 4 Conclusion

While this paper cannot prove the existence of an aether, it has been shown that the two most iconic features of the theory of special relativity, the mass-energy equivalency (eq. 6) and the Lorentz factor (eq 8), are identical to the expressions for compression of a fluid with the speed of sound equal to the speed of light in free space. Is it then most reasonable to assume that they are due to

   a) compression of a fluid?
   b) curvature of space and time itself?

It is at least to me very difficult to prefer option b) to option a). It should also be remembered that there has never been any proof against the existence of the aether and that the SR took preference over the Lorentz aether theory on the belief that it was based on less arbitrary rules. The aether was abandoned because it was not needed for the SR and this works mathematically but has left us with a very surreal world view from a physical sense.

Finally, I believe the reintroduction of the aether makes modern physics infinitely more palatable: Relativistic effects in both special and general relativity would be due to compression of actual matter rather than compression of time and space itself and quantum physics now would describe waves in real matter rather than abstract probability functions in free space, in a similar way as Erwin Schrödinger originally interpreted the wave function in his famous equation as a charge distribution in space [28]. Wave-particle dualism of e.g. electrons would become much easier to accept if there is something tangible oscillating. Certain aspects of classical physics also would become a lot more palatable as the aether provides a medium to transmit otherwise very strange distant forces from gravitational and electrical fields.

## Acknowledgements

I would like to thank Markus Gerdin, Ph.D. and Piotr Migdal, Ph.D. for valuable discussion of this paper.

## References

[1]   C. Huygens, *Traité de la lumière*. 1690.
[2]   I. Newton, "Hypothesis explaining the properties of light," ed, 1675.
[3]   T. Young, "Bakerian Lecture: Experiments and calculations relative to physical optics," *Philosophical Transactions of the Royal Society,* vol. 94, pp. 1-16, 1804.
[4]   F. Arago and A.-J. Fresnel, "On the Action of Rays of Polarized Light upon Each Other," *Annales de Chimie et de Physique,* 1819.
[5]   J. C. Maxwell, "On physical lines of force," *The London, Edinburgh and Dublin philosophical magazine and journal of science,* 1861.
[6]   G. G. Stokes, "On the aberration of light," *Philosophical Magazine,* vol. 27, no. 177, pp. 9-15, 1845.
[7]   A. A. Michelson, "The Relative Motion of the Earth and the Luminiferous Ether," *American Journal of Science,* vol. 22, pp. 120-129, 1881.
[8]   A. A. Michelson and E. Morley, "On the Relative Motion of the Earth and the Luminiferous Ether," *American Journal of Science,* vol. 34, pp. 333-345, 1887.
[9]   H. Fizeau, " The Hypotheses Relating to the Luminous Aether, and an Experiment which Appears to Demonstrate that the Motion of Bodies Alters the Velocity with which Light Propagates itself in their Interior," *Philosophical Magazine, Series 4,* vol. 2, pp. 568-573, 1851.
[10]  G. F. FitzGerald, "The Ether and the Earth's Atmosphere," *Science,* vol. 13, no. 328, p. 390, 1889.
[11]  H. A. Lorentz, *Versuch einer Theorie der electrischen und optischen Erscheinungen in bewegten Körpern*. Leiden: E.J. Brill, 1895.
[12]  J. D. Van de Ven and P. Y. Li, "Liquid piston gas compression," (in English), *Appl Energ,* vol. 86, no. 10, pp. 2183-2191, Oct 2009.
[13]  A. Einstein, "Ist die Trägheit eines Körpers von seinem Energieinhalt abhängig?," *Annalen der Physik,* vol. 18, pp. 639-643, 1905.
[14]  H. Minkowski, "Raum un Zeit," *Physikalische Zeitschrift,* vol. 10, pp. 75-88, 1909.
[15]  A. Einstein, "Zur Elektrodynamik bewegter Körper " *Annalen der Physik (ser. 4),* vol. 17, pp. 891-921, 1905.
[16]  H. Glauert, "The Effect of Compressibility on the Lift of an Aerofoil " *Proceedings of the Royal Society of London. Series A, Containing Papers of a Mathematical and Physical Character,* vol. 118, no. 779, pp. 113-119, 1928.
[17]  S. W. Rienstra and A. Hirschberg, *An Introduction to Acoustics*, Eindhoven University of Technology, 2016. [Online]. Available: https://www.win.tue.nl/~sjoerdr/papers/boek.pdf.
[18]  A. Nishida, "On the Singularity at Light Velocity," 2013 [Online]. Available: http://www.ssisc.org/people/anishida/files/papera.pdf
[19]  A. Einstein, "Über den Äther," *Verhandlungen der Schweizerischen naturforschenden Gesellschaft,* vol. 105 (pt. 2), pp. 85-93, 1924.
[20]  P. Dirac, "Is there an Æther?," *Nature,* no. 168, pp. 906-907, 1951.
[21]  Y. Lee, T. M. Haard, W. P. Halperin, and J. A. Sauls, "Discovery of the acoustic Faraday effect in superfluid He-3-B," (in English), *Nature,* vol. 400, no. 6743, pp. 431-433, Jul 29 1999.
[22]  K. P. Sinha, C. Sivaram, and E. C. G. Sudarshan, "Aether as a Superfluid State of Particle-Antiparticle Pairs," (in English), *Found Phys,* vol. 6, no. 1, pp. 65-70, 1976.

[23] J. C. Kapteyn, "First attempt at a theory of the arrangement and motion of the sidereal system," *Astrophysical Journal,* vol. 55, pp. 302-327, 1922.

[24] W. Thomson, "On vortex atoms," *Proceedings of the Royal Society of Edinburgh,* vol. 6, pp. 94-105, 1867.

[25] E. J. Aiton, "The Cartesian theory of gravity," *Annals of Science,* vol. 15, no. 1, pp. 27-49, 1959.

[26] W. Bertozzi, "Speed and Kinetic Energy of Relativistic Electrons " *American Journal of Physics,* vol. 32, no. 7, pp. 551-555, 1964.

[27] J. C. Hafele and R. E. Keating, "Around-the-World Atomic Clocks: Predicted Relativistic Time Gains," *Science,* vol. 177, no. 4044, pp. 166-170, 1972.

[28] E. Schrödinger, "An Undulatory Theory of the Mechanics of Atoms and Molecules," *Physical Review,* vol. 28, no. 6, pp. 1049-1070, 1926.